\documentclass[apj]{emulateapj}

\usepackage{threeparttable}

\shorttitle{Supernova Search at Super-Kamiokande}
\shortauthors{Ikeda et al. (Super-Kamiokande Collaboration)}

\begin{document}

\title{Search for Supernova Neutrino Bursts at Super-Kamiokande}

\newcounter{foots}
\newcounter{notes}
\newcommand{\authoraticrr}{$^{1}$}
\newcommand{\authoratncen}{$^{2}$}
\newcommand{\authoratbu}{$^{3}$}
\newcommand{\authoratbupenn}{$^{3,\dagger}$}
\newcommand{\authoratbnl}{$^{4}$}
\newcommand{\authoratuci}{$^{5}$}
\newcommand{\authoratcsu}{$^{6}$}
\newcommand{\authoratcnu}{$^{7}$}
\newcommand{\authoratduke}{$^{8}$}
\newcommand{\authoratgmu}{$^{9}$}
\newcommand{\authoratgifu}{$^{10}$}
\newcommand{\authoratuh}{$^{11}$}
\newcommand{\authoratui}{$^{12}$}
\newcommand{\authoratkek}{$^{13}$}
\newcommand{\authoratkekicrr}{$^{13,1}$}
\newcommand{\authoratkekkashiwa}{$^{13,\ddagger}$}
\newcommand{\authoratkobe}{$^{14}$}
\newcommand{\authoratkyoto}{$^{15}$}
\newcommand{\authoratkyototriumf}{$^{15,\S}$}
\newcommand{\authoratlanluci}{$^{16,5}$}
\newcommand{\authoratlsu}{$^{17}$}
\newcommand{\authoratumd}{$^{18}$}
\newcommand{\authoratduluth}{$^{19}$}
\newcommand{\authoratmiyagi}{$^{20}$}
\newcommand{\authoratnagoya}{$^{21}$}
\newcommand{\authoratsuny}{$^{22}$}
\newcommand{\authoratniigata}{$^{23}$}
\newcommand{\authoratokayama}{$^{24}$}
\newcommand{\authoratosaka}{$^{25}$}
\newcommand{\authoratseoul}{$^{26}$}
\newcommand{\authoratshizuoka}{$^{27}$}
\newcommand{\authoratshizuokaseika}{$^{28}$}
\newcommand{\authoratskku}{$^{29}$}
\newcommand{\authorattohoku}{$^{30}$}
\newcommand{\authorattokai}{$^{31}$}
\newcommand{\authorattit}{$^{32}$}
\newcommand{\authorattokyo}{$^{33}$}
\newcommand{\authorattsinghua}{$^{34}$}
\newcommand{\authoratwarsaw}{$^{35}$}
\newcommand{\authoratwarsawuci}{$^{35,5}$}
\newcommand{\authoratuw}{$^{36}$}
\newcommand{\authoratuwduluth}{$^{36,19}$}
\newcommand{\addressoficrr}[1]{$^{1}$ #1 }
\newcommand{\addressofncen}[1]{$^{2}$ #1 }
\newcommand{\addressofbu}[1]{$^{3}$ #1 }
\newcommand{\addressofbnl}[1]{$^{4}$ #1 }
\newcommand{\addressofuci}[1]{$^{5}$ #1 }
\newcommand{\addressofcsu}[1]{$^{6}$ #1 }
\newcommand{\addressofcnu}[1]{$^{7}$ #1 }
\newcommand{\addressofduke}[1]{$^{8}$ #1 }
\newcommand{\addressofgmu}[1]{$^{9}$ #1 }
\newcommand{\addressofgifu}[1]{$^{10}$ #1 }
\newcommand{\addressofuh}[1]{$^{11}$ #1 }
\newcommand{\addressofui}[1]{$^{12}$ #1 }
\newcommand{\addressofkek}[1]{$^{13}$ #1 }
\newcommand{\addressofkobe}[1]{$^{14}$ #1 }
\newcommand{\addressofkyoto}[1]{$^{15}$ #1 }
\newcommand{\addressoflanl}[1]{$^{16}$ #1 }
\newcommand{\addressoflsu}[1]{$^{17}$ #1 }
\newcommand{\addressofumd}[1]{$^{18}$ #1 }
\newcommand{\addressofduluth}[1]{$^{19}$ #1 }
\newcommand{\addressofmiyagi}[1]{$^{20}$ #1 }
\newcommand{\addressofnagoya}[1]{$^{21}$ #1 }
\newcommand{\addressofsuny}[1]{$^{22}$ #1 }
\newcommand{\addressofniigata}[1]{$^{23}$ #1 }
\newcommand{\addressofokayama}[1]{$^{24}$ #1 }
\newcommand{\addressofosaka}[1]{$^{25}$ #1 }
\newcommand{\addressofseoul}[1]{$^{26}$ #1 }
\newcommand{\addressofshizuoka}[1]{$^{27}$ #1 }
\newcommand{\addressofshizuokaseika}[1]{$^{28}$ #1 }
\newcommand{\addressofskku}[1]{$^{29}$ #1 }
\newcommand{\addressoftohoku}[1]{$^{30}$ #1 }
\newcommand{\addressoftokai}[1]{$^{31}$ #1 }
\newcommand{\addressoftit}[1]{$^{32}$ #1 }
\newcommand{\addressoftokyo}[1]{$^{33}$ #1 }
\newcommand{\addressoftsinghua}[1]{$^{34}$ #1 }
\newcommand{\addressofwarsaw}[1]{$^{35}$ #1 }
\newcommand{\addressofuw}[1]{$^{36}$ #1 }

\def\pennnow{$\dagger$}
\def\kashiwanow{$\ddagger$}
\def\triumfnow{\S}
\author{
{\bf The Super-Kamiokande Collaboration} \\
\vspace{0.2cm}
M.~Ikeda\authoratokayama,
A.~Takeda\authoraticrr,
Y.~Fukuda\authoratmiyagi,
M.R.~Vagins\authoratuci,
K.~Abe\authoraticrr,
T.~Iida\authoraticrr,
K.~Ishihara\authoraticrr,
J.~Kameda\authoraticrr,
Y.~Koshio\authoraticrr,
A.~Minamino\authoraticrr
% ...
\\
\smallskip
\smallskip
\footnotesize
\it
\addressoficrr{Kamioka Observatory, Institute for Cosmic Ray Research, 
University of Tokyo, Kamioka, Gifu, 506-1205, Japan}\\
\addressofncen{Research Center for Cosmic Neutrinos, Institute for Cosmic 
Ray Research, University of Tokyo, Kashiwa, Chiba 277-8582, Japan}\\
\addressofbu{Department of Physics, Boston University, Boston, MA 02215, 
USA}\\
\addressofbnl{Physics Department, Brookhaven National Laboratory, Upton, 
NY 11973, USA}\\
\addressofuci{Department of Physics and Astronomy, University of 
California, Irvine, Irvine, CA 92697-4575, USA }\\
\addressofcsu{Department of Physics, California State University, 
Dominguez Hills, Carson, CA 90747, USA}\\
\addressofcnu{Department of Physics, Chonnam National University, Kwangju 
500-757, Korea}\\
\addressofduke{Department of Physics, Duke University, Durham, NC 27708, 
USA} \\
\addressofgmu{Department of Physics, George Mason University, Fairfax, VA 
22030, USA }\\
\addressofgifu{Department of Physics, Gifu University, Gifu, Gifu 
501-1193, Japan}\\
\addressofuh{Department of Physics and Astronomy, University of Hawaii, 
Honolulu, HI 96822, USA}\\
\addressofui{Department of Physics, Indiana University, Bloomington,
  IN 47405-7105, USA} \\
\addressofkek{High Energy Accelerator Research Organization (KEK), 
Tsukuba, Ibaraki 305-0801, Japan }\\
\addressofkobe{Department of Physics, Kobe University, Kobe, Hyogo 
657-8501, Japan}\\
\addressofkyoto{Department of Physics, Kyoto University, Kyoto 606-8502, 
Japan}\\
\addressoflanl{Physics Division, P-23, Los Alamos National Laboratory, Los 
Alamos, NM 87544, USA }\\
\addressoflsu{Department of Physics and Astronomy, Louisiana State 
University, Baton Rouge, LA 70803, USA }\\
\addressofumd{Department of Physics, University of Maryland, College Park, 
MD 20742, USA }\\
\addressofduluth{Department of Physics, University of Minnesota, Duluth, 
MN 55812-2496, USA}\\
\addressofmiyagi{Department of Physics, Miyagi University of Education, 
Sendai, Miyagi 980-0845, Japan}\\
\addressofnagoya{Solar Terrestrial Environment Laboratory, Nagoya University, Nagoya, Aichi 
464-8602, Japan}\\
\addressofsuny{Department of Physics and Astronomy, State University of 
New York, Stony Brook, NY 11794-3800, USA}\\
\addressofniigata{Department of Physics, Niigata University, Niigata, 
Niigata 950-2181, Japan }\\
\addressofokayama{Department of Physics, Okayama University, Okayama, 
Okayama 700-8530, Japan} \\
\addressofosaka{Department of Physics, Osaka University, Toyonaka, Osaka 
560-0043, Japan}\\
\addressofseoul{Department of Physics, Seoul National University, Seoul 
151-742, Korea}\\
\addressofshizuoka{Department of Systems Engineering, Shizuoka University, 
Hamamatsu, Shizuoka 432-8561, Japan}\\
\addressofshizuokaseika{Department of Informatics in Social Welfare, Shizuoka University 
of Welfare, Yaizu, Shizuoka, 425-8611, Japan}\\
\addressofskku{Department of Physics, Sungkyunkwan University, Suwon 
440-746, Korea}\\
\addressoftohoku{Research Center for Neutrino Science, Tohoku University, 
Sendai, Miyagi 980-8578, Japan}\\
\addressoftokai{Department of Physics, Tokai University, Hiratsuka, 
Kanagawa 259-1292, Japan}\\
\addressoftit{Department of Physics, Tokyo Institute for Technology, 
Meguro, Tokyo 152-8551, Japan }\\
\addressoftokyo{The University of Tokyo, Tokyo 113-0033, Japan }\\
\addressoftsinghua{Department of Engineering Physics, Tsinghua University, Beijing, 100084, China}\\
\addressofwarsaw{Institute of Experimental Physics, Warsaw University, 
00-681 Warsaw, Poland }\\
\addressofuw{Department of Physics, University of Washington, Seattle, WA 
98195-1560, USA}
}
\altaffiltext{\pennnow}{Present address: Center for Gravitational Wave Physics, Pennsylvania State 
University, University Park, PA 16802, USA}

\begin{abstract}
The result of a search for neutrino bursts from supernova explosions 
using the Super-Kamiokande detector is reported.
Super-Kamiokande is sensitive to core-collapse supernova explosions
via observation of their neutrino emissions. 
The expected number of events comprising such a burst is $\sim 10^4$
and the average energy of the neutrinos is in few tens of MeV range
in the case of a core-collapse supernova explosion at the typical distance in our galaxy (10 kiloparsecs);
this large signal means that the detection efficiency anywhere within our galaxy and well
past the Magellanic Clouds is 100\%.  
We examined a data set which was taken from May, 1996 to July, 2001
and from December, 2002 to October, 2005 
corresponding to 2589.2 live days.
However, there is no evidence of such a supernova explosion during the data-taking period.
The 90\% C.L. upper limit on the rate of core-collapse supernova explosions out to distances of
100 kiloparsecs is found to be 0.32 SN $\cdot $ year$^{-1}$. 
\end{abstract}

\keywords{galaxies: individual (our Galaxy,LMC,SMC) --- neutrinos --- supernovae: general}

\section{Introduction}
	On the 23rd of February, 1987, the Kamiokande II,
IMB, and Baksan experiments observed the neutrino burst from SN1987A, which
was located in the Large Magellanic Cloud ~\cite{KAM,IMB,BKS}
\footnote{The LSD detector observed a cluster of 5 events about 5 hours earlier,
but the correlation between this signal and SN1987A is favored only by  
non-standard double-bang scenarios of stellar collapse \cite{LSD}.}.  
This was the first detection of a supernova's neutrino burst, 
and it introduced a new method of investigation: neutrino astronomy.
	
	Super-Kamiokande (Super-K, SK) is an imaging water Cherenkov detector
containing 50,000 tons of pure water; it is the
successor to the Kamiokande detector. Super-K is located 1000 meters
underground (2,700 meters of water equivalent) in the Kamioka zinc mine in
the Gifu prefecture of Japan, at 36.4$^{\circ}$N, 137.3$^{\circ}$E and
25.8$^{\circ}$N geomagnetic latitude. The detector consists of a main
inner detector and an outer veto detector. Both detectors are contained within a
cylindrical stainless steel tank 39.3 m in diameter $\times$ 41.4 m in
height.  The usual fiducial mass for neutrino measurements is 22.5 ktons with
boundaries 2.0 m from the inner surface. The outer detector is also 
a water Cherenkov detector of 13,000 metric tons total mass.  It 
surrounds the inner detector as a 4$\pi$ solid-angle anti-detector to
detect any signals coming from outside of the detector and to shield
against external gamma-rays and neutrons.
	 
        The data set for the analyses was taken during two periods.
The first period started on the
1st of April, 1996, and terminated on the 15th of July, 2001. 
A total of 11,146 photo multipliers (PMT's) with 20-inch diameter  
photocathodes provided active light collection over 40\% of the 
entire surface of the inner detector. 
This phase of the project is now referred to as Super-Kamiokande-I (SK-I). 
The second phase, Super-Kamiokande-II (SK-II), started on 
the 10th of December, 2002, and terminated on the 6th of October, 2005.
A total of 5,182 20-inch PMT's, each protected by acrylic and fiber-reinforced plastic 
(FRP) cases, were mounted on the inner detector, providing 19\% photocathode coverage 
during this period.
In the outer detector, a total of 1,885 8-inch PMT's were installed during 
both periods. 
Due to rapid variations in the water transparency, the data before the 31st of May,
1996 in SK-I, and the data before the 23rd of December, 2002 in SK-II
have not been used for analysis because of uncertainties in the energy calibration.
Having a well-defined energy response is necessary when searching for low 
multiplicity event clusters, as will be described in the later sections of this paper.  

	Theoretical calculations predict the characteristics of 
neutrinos expected from a supernova. 
A typical core-collapse supernova explosion emits all types of neutrinos
and has a total energy output of $\sim3\times 10^{53}$ ergs which 
would generate about 10,000 SK events (9,000 events without neutrino oscillation) 
in the case of a supernova at a distance of 10 kpc from the earth.
Neutrino oscillations enhance the
overall flux we observe by 10\% as higher-temperature 
$\nu_{\mu}(\overline{\nu}_{\mu})$ and $\nu_{\tau}(\overline{\nu}_{\tau})$
get fully mixed into the observable $\nu_{e}(\overline{\nu}_{e})$ signal.

        These supernova events will be detected via the following interactions in SK,
where the numbers in parentheses show the fractions of the total number of events 
with/without neutrino oscillations,
\begin{eqnarray}
&\overline{\nu}_e + p        \rightarrow  n + e^+ &(88\%/89\%)\:, \label{nebp}\\
&\nu_e +  e^-                \rightarrow  \nu_e + e^- &(1.5\%/1.5\%)\:,\label{nee}\\
&\overline{\nu}_e +  e^-     \rightarrow  \overline{\nu}_e + e^- &(<1\%/<1\%)\:,\\
&\nu_x +e^-                  \rightarrow  \nu_x + e^- &(1\%/1\%)\:,\\
&\nu_e +  ^{16}O             \rightarrow  e^{-} + ^{16}F  &(2.5\%/<1\%)\:,\label{cc1}\\
&\overline{\nu}_e +  ^{16}O  \rightarrow  e^{+} + ^{16}N &(1.5\%/1\%)\:,and \label{cc2}\\
&\nu_x +^{16}O               \rightarrow \nu_x +O^{*}/N^{*}+\gamma &(5\%/6\%)\:,\label{nc}
\end{eqnarray}
and $\nu_x$ means the total interactions of $\nu_{\mu}$,$\nu_{\tau}$, and
their anti-neutrinos. To obtain these fractions \cite{OSC}, a supernova neutrino burst using 
the Livermore model \cite{Totani} is assumed.
To account for neutrino oscillations in Ref. (Takahashi et al.~2001),
$\theta_{12}$ and $\Delta m_{12}^{2}$ are set 
to be in the favored solar neutrino LMA region, 
and $\theta_{23}$ and $\Delta m_{23}^{3}$ are determined from the best oscillation 
fit for the atmospheric neutrinos.
For the fraction of the neutral current interactions with oxygen
(Eq.~\ref{nc}),
we look to Ref.\cite{langanke} and obtain our number
by scaling their result to the numbers in Ref. \cite{OSC}.
At present, based on both data and models, many of these numbers
are uncertain up to a factor of two, especially the numbers of
interactions with oxygen. In particular, Eqs.\ref{cc1}, \ref{cc2}, and \ref{nc}
are very dependent on the neutrino temperatures and details of mixing
as discussed in the cited literature.

        Averaged energies of neutrinos from a standard delayed explosion supernova 
model are expected to be between 11 $\sim$ 26 MeV 
(Totani et al.~1998; Thompson et al.~2003; Sumiyoshi et al.~2005) 
For example, the calculation performed by the Livermore group
shows  $\langle E_{\nu_e} \rangle \sim 11$ MeV,
$\langle E_{\overline{\nu}_e}  \rangle \sim 16$ MeV and 
$\langle E_{\nu_{\mu/\tau}} \rangle \sim25$ MeV ~\cite{Totani}.
The time profile of each type of neutrino has a unique shape. 
During the initial $\sim$10 milliseconds, electron neutrinos from the neutronization are
released with a total energy on the order of $10^{51}$ ergs. After the
neutronization, all flavors of neutrinos are produced by electron-positron
annihilation, and released with a total energy on the order of $10^{53}$ ergs
with a time scale of several tens of seconds. 

	It still is not clear at present, however, whether or not the  
delayed explosion scenario is the correct picture, in which
a stalled shock wave will be reheated by energy deposition from neutrinos.
We expect that the observation of a galactic supernova
by the SK detector with high statistics will solve this long-standing 
question and, furthermore, shed light on other important questions
such as neutrino oscillations and neutrino mass ~\cite{Raffelt}.

In this paper, we have searched for supernova neutrino bursts using 
the SK detector, and we present  
an upper limit on the supernova explosion rate 
within 100 kpc of the earth (this volume contains our Galaxy, the LMC, and the SMC).
We also report the first result of search for neutronization bursts. 
Previously, AMANDA, Baksan, IMB, and MACRO have reported upper limits on 
the supernova explosion rate just within our Galaxy 
\cite{AMND,BKS_sn,IMB_sn,MCR}, while LVD and Kamiokande have reported 
preliminary results \cite{LVD,kam} of the same nature.
 
\section{Data analysis}

\subsection{Data set}

\begin{figure}
\epsscale{1.0}
\plotone{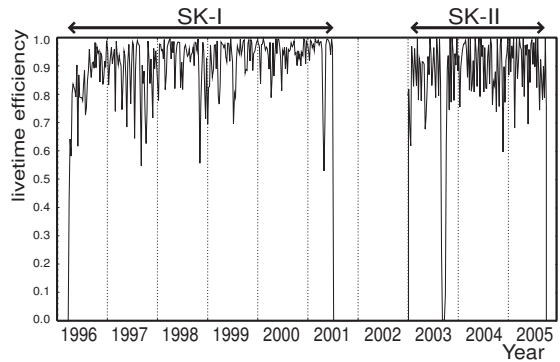}
\caption{SK livetime efficiency as a function of time. 
Most of the big dips in efficiency were due to planned calibration runs such as LINAC
energy calibration, though the biggest dip in 2003 was due 
to electronics problems.\label{livtim}}
\end{figure}

	From the 31st of May, 1996, to the 15th of July, 2001,
and from the 24th of December, 2002, to the 5th of October, 2005, the 
livetimes of our detector for supernova searches were 1703.9 days for SK-I,
and 885.3 days for SK-II. Livetime efficiency as a
function of date is shown in Figure~\ref{livtim} --- over the course of
the entire data-taking period the average efficiency was about 89\%. 
The primary cause of lost supernova livetime was scheduled calibration work.
The large losses of livetime seen in Figure~\ref{livtim} around
July 1997, December 1998, June 1999, May 2001, March 2003, and September 2005
were due to LINAC energy calibration runs~\cite{LINAC}, though the biggest 
dip in 2003 was due to electronics problems which have since been corrected.

       Vertex and energy reconstruction techniques are the same as those used in
our solar neutrino analysis~\cite{SK}. Fiducial volume for the supernova
search is also 22.5 ktons, though the energy thresholds are 6.5~MeV (SK-I) 
and 7.0 MeV (SK-II) to avoid the higher background rates associated with 
the lower thresholds used in the solar neutrino analysis.
Data reduction steps are basically the same. The first reduction includes removing 
events due to electronics noise, events with a poor reconstruction
of the vertex position, and events with a vertex outside of the fiducial
volume. A spallation cut removes events which are produced by energetic
muons by using a likelihood method in which both the time difference
$\Delta T$ and distance $\Delta L$ between the parent muon and subsequent
events are considered~\cite{SK}. Mis-reconstructed events are also
removed. 
After these noise reductions, the remaining event rates 
are 180 events/day for SK-I, and 164 events/day for SK-II. 

\subsection{Analysis method}

	In this section, we introduce our general method of 
supernova burst search. The procedure of data analysis is as follows:
\begin{enumerate}
 \item Scan event times in the data set using a sliding time-window.
 We define a ``cluster'' 
 if the number of events (the ``multiplicity'') within a time-window
 is greater or equal to a certain threshold. 
 \item Check each cluster found in the first step to determine if 
 it is a real signal from a supernova or the result of background events.

\end{enumerate}

         A background cluster consists of either time-correlated 
non-supernova events or a chance coincidence of uncorrelated low energy events.
There are two dominant background sources which produce time-correlated events.
One such source is flasher PMT's which act as sources of light, 
and the other is spallation product, radioactive isotopes 
made via interactions between energetic cosmic ray muons and oxygen nuclei.
In both cases, the reconstructed vertices of 
the resulting events are spatially concentrated.
On the other hand, events made by actual supernova neutrinos  
should be generated uniformly in the detector volume.
Therefore, to distinguish real signals from background clusters, 
clusters are checked by studying the correlation between the
multiplicity and the events' spatial distribution ($R_{mean}$).
$R_{mean}$ is defined by the
averaged spatial distance between each event as follows:
\begin{equation}
 R_{mean} = \frac{\displaystyle{\sum_{i = 1}^{M - 1}} \;
 \displaystyle{\sum_{j = i + 1}^{M}} \vert \vec{r_i} -
\vec{r_j} \vert}{_{M}C_2} .
\end{equation}
where $M$ is the multiplicity, $\vert \vec{r_i} - \vec{r_j} \vert$
is the distance between events $i$ and $j$, and $_{M}C_2$ is the
number of unique combinations.  In the case of a supernova
burst, $R_{mean}$ should have a larger value than that resulting from
spatially clustered events such as spallation events or flasher events.

\begin{figure}
\epsscale{1.0}
\plotone{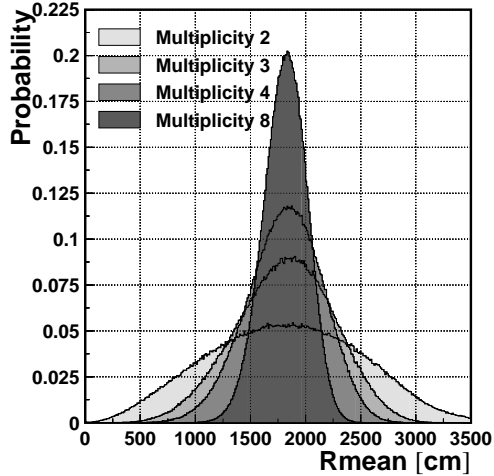}
\caption{Expected $R_{mean}$ distributions of supernova events
obtained by Monte Carlo simulation 
for multiplicity equal to 2, 3, 4, and 8.\label{rmean_mc}}
\end{figure}

          Figure~\ref{rmean_mc} shows $R_{mean}$
distributions of simulated supernova events for 
multiplicity 2, 3, 4, and 8, which are the 
multiplicity thresholds of various burst searches discussed
in later sections of this paper.
As shown in this figure, if the events in a cluster occur
uniformly in the detector, as in supernova, 
then the expected value of $R_{mean}$ tends to be
large, around 1,800 cm. The threshold of $R_{mean}$
for a cluster with multiplicity equal to 2 is set to 750 cm,
and for clusters  with multiplicity greater than 2 
the threshold is set to 1,000 cm.
The efficiencies for supernova events using these criteria
are 94\%, 96\%, 99\%, and 100\%, respectively.

In the following sections, 
we obtain results from three kinds of burst searches
by changing the width of the time-window and the multiplicity threshold 
and setting additional criteria for respective searches.
The three burst searches are 
1) distant supernova search, 2) supernova burst search with low energy threshold,
and 3) neutronization burst search.

\subsection{Distant supernova search}
 
	In recent years, a lot of effort has been put into optical supernova searches,
with the result that the number of detected supernovae has increased substantially. 
The optical detection rate of supernovae suggests that theoretical
calculations of supernova rates might be
significantly underestimated. More specifically, nine core-collapse supernovae 
have been found in nearby galaxies between 2002 and 2005; 
this is three times higher than one theoretical estimate~\cite{ando}.
 
	This motivates us to search for neutrinos 
from supernovae in nearby galaxies. 
Since the expected total number of events at SK from a supernova in the 
Andromeda galaxy ($\sim $700kpc) is around two events, 
for this search we set our criteria as a long time-window 
and low multiplicity threshold such as $\geq$ 2 events / 20 seconds. 
A lot of background clusters would be found in our usual 
solar neutrino data set~\cite{SK} with this criterion.
Therefore, we reduce those backgrounds by setting a higher energy threshold
because the average energy of the emitted positron from 
the interaction (Eq.\ref{nebp}) is higher than that of most low energy 
background events.
To find the best energy threshold, 
we calculate the detection probability of a supernova
and the number of chance coincidences of the low energy background events
as functions of energy threshold value. 
Then we take the value of energy threshold which gives us the maximum value of 
$ \frac{\text{Detection probability}}{\sqrt{\# \text{of chance coincidences}}}$
as an optimized energy threshold. The resulting value we set is 17 MeV,
where the single event rate with this energy threshold is 0.762 event/day for SK-I,
and 1.03 event/day for SK-II.

\begin{figure}[h]
\epsscale{1.0}
\plotone{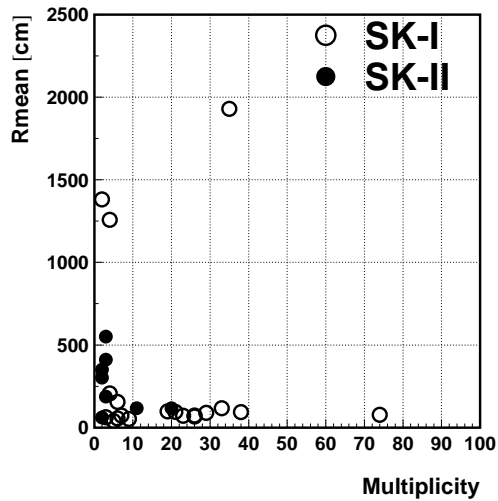}
\caption{Correlation between the multiplicity and $R_{mean}$
for obtained clusters from data (open circle: SK-I, closed circle: SK-II ).
There were 19 clusters observed in SK-I and 8 clusters 
in SK-II.\label{ltw}}
\end{figure}

	Figure~\ref{ltw} is a scatter plot of $R_{mean}$ versus the multiplicity
for each cluster which satisfies the criteria with an 
energy threshold of 17 MeV. Three candidate clusters exist that have $R_{mean}$ more than 
1000 cm, 
but as shown in Table~\ref{time_table},
the event times of those candidates actually coincide with times of mine blasting.
Because of the physical vibration of PMT's due to blasting, 
huge electrical noise occurred, 
and those noises caused time-clustered 
a few hundreds events within a few seconds during or after blasting.
It was also confirmed that these events including the candidate cluster events
have a characteristic PMT hit pattern
naerly in the same area of the inner detector, and hence they should be 
eliminated as potential supernova events.

\begin{deluxetable}{cccc}
\tablewidth{0pc}
\tablecaption{Detection time of candidates in SK-I \label{time_table}}
\tablehead{\colhead{Candidate}    &  \colhead{Date and Time (JST)\tablenotemark{a}} &   \colhead{Rmean}   & \colhead{Multiplicity}}
\startdata
 No.1	&	Jul.13th 1999 19:00	&	1257.89	&	4\\
 No.2	&	May.12th 2000 11:06	&	1928.76	&	35\\
 No.3	&	Oct.12th 2000 19:03	&	1381.76 &	2\\
\enddata
\tablenotetext{a}{The blastings by Kamioka Mining and Smelting Company
 were scheduled at 11:00, 19:00, and 23:00, and shift takers reported
 that there were blastings at those times.}
\end{deluxetable}

The background clusters which have small $R_{mean}$ values in Fig.~\ref{ltw}
were mainly found around times of calibration work,
and more importantly the positions of these clusters were consistent with 
the positions of the calibration sources themselves.  That these clusters made it 
into our supposedly all non-calibration data sample is therefore judged to be 
the result of occasional operator error in assigning the proper run type during 
transitions between normal and calibration data taking.
(Note that such mis-labeled runs are excluded in our usual solar neutrino data analysis,
but in order to maximize supernova livetime a wider variety of runs were included 
in this supernova analysis.)
Therefore, no real supernova signal was observed during 
the data-taking periods in both SK-I and SK-II.  

\subsection{Supernova burst search \\ with low energy threshold}

	In this section, we set various time-windows 
of 0.5, 2, and 10 seconds with a lower energy threshold of 
6.5 MeV for SK-I and 7.0 MeV for SK-II
to search for signals from a supernova
in this lower energy region. 
In fact, the data of SN1987A imply $\langle E  \rangle = 7.5 $ MeV
at Kamiokande and 11.1 MeV at IMB \cite{Jegerlehner}
while $\langle E_{\overline{\nu}_e}  \rangle \sim 16$ MeV
to take the Livermore group model as an example \cite{Totani}.

	Since a lot of background events due to spallation or flasher PMT's
still remain in the lower energy regions,
we set criteria of higher multiplicity for each time-window as 
$\geq$ 3 events / 0.5 seconds, $\geq$ 4 events / 2.0 seconds, 
and $\geq$ 8 events/ 10 seconds.
If a cluster satisfies the requirement $R_{mean}$,
we closely check data around the cluster --- 
for example all events within $\pm$20 seconds around the cluster --- 
so that we get information from as many supernova neutrinos as possible.  

\begin{figure}[h]
\epsscale{1.0}
\plotone{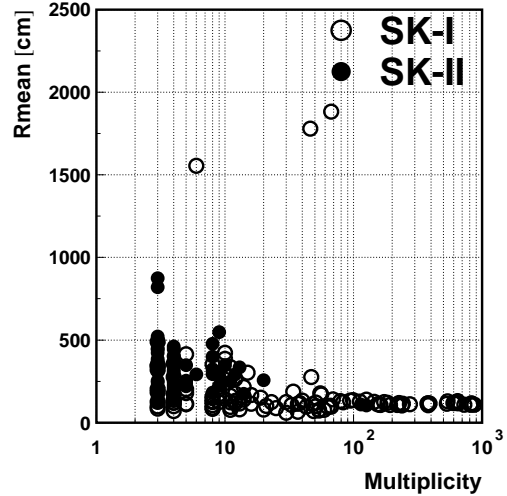}
\caption{Correlation between the multiplicity and $R_{mean}$
for obtained clusters from data (open circle: SK-I, closed circle: SK-II).
There are 121 and 53 clusters observed in SK-I 
and SK-II respectively. If a cluster satisfies more than one criterion,
the circle for the cluster represents the largest multiplicity, and
$R_{mean}$ for the cluster is calculated from the multiplicity of events.
\label{ml_rm_data}}
\end{figure}

	Figure~\ref{ml_rm_data} shows the correlation between $R_{mean}$ 
and multiplicity for obtained clusters from SK-I and SK-II.
There were only three clusters which had large $R_{mean}$ values in SK-I.
As mentioned in the previous section, however, all three candidate clusters
consist of the same events as in the previous section which 
were found during periods of mine blasting.
 
	The same criteria are applied to SK-II data, and the correlation of
the multiplicity and $R_{mean}$ of the candidate clusters was studied.
There was no cluster with $R_{mean} \geq $ 1000~cm,
but two candidate clusters had $R_{mean}$ just below 1000~cm. 
To make sure that those candidates were not real signals,
cosmic ray muon events around the cluster were studied, and
event displays of candidate events were checked.
As a result, one of the candidates was found to be an accidental coincidence of 
a low energy event with spallation events.
Although most spallation events have already been removed by the
spallation cut, a small fraction of spallation products are still present
at lower multiplicities. 
The other candidate cluster was comprised of false events caused by a flasher tube.
As in the previous search, it was confirmed that 
the background clusters with small $R_{mean}$
were mostly due to mislabeled calibration work. 

	In conclusion, the remaining candidate clusters were all caused by
mine blasting, and so there is no clear evidence for any supernova
neutrino burst in the data obtained by SK-I as well as SK-II.
Therefore, we will present an upper limit at the 90\% confidence level for 
the rate of supernova explosions in Section 3.

\subsection{Neutronization burst search} 

	We conducted another burst search with shorter time-windows
to investigate the short-lived 
neutronization burst of $\nu_e$ events from a supernova.
Prior to the core explosion, many $\nu_e$'s are emitted via the reaction 
$e^- + p \rightarrow \nu_e + n$ as the shock wave propagates into the exploding star's  
outer core. The shock wave dissociates nuclei into free nucleons
on which the cross section of electron capture is larger than that on nuclei; 
the resulting burst of $\nu_e$'s thus forms the so-called neutronization burst.

	The duration of the neutronization burst is on the time scale of 
the shock wave propagation, which is less than 10 milliseconds. 
Hence, if such a short time window is set, the $\nu_e$'s can be dominant, 
and the expected number of neutronization burst neutrinos which will be observed at 
SK is between one and six (depending on neutrino oscillation models)
in the case of a supernova at a distance of 10 kpc from the earth~\cite{OSC}.
It should be mentioned that even if no supernova explosion is observed in the galaxy
during that period, it still might be possible to observe only the neutronization burst.
For example, if a black hole forms shortly after the neutronization stage,
then the main burst of supernova neutrinos might not be able to escape 
from the black hole~\cite{sumi}.

	Based on the above theoretical expectations, 
we set time-windows of 1, 10, and 100 milliseconds. 
The multiplicity threshold is two events for each time window;
$\geq$ 2 events / 1 msec, $\geq$ 2 events / 10 msec, or $\geq$ 2 events / 100 msec.
Because recoil electrons will have lower 
energies from neutrino--electron scattering (Eq.\ref{nee})
which is the dominant interaction in this case,
we use the same sample as the SK-I solar neutrino analysis
\cite{SK}, whose energy threshold is 5 MeV and livetime is 1496 days,
and we use the same sample as in previous sections for SK-II.

\begin{figure}[h]
\epsscale{1.0}
\plotone{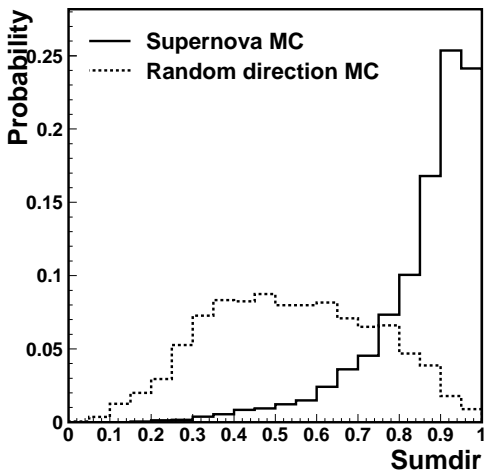}
\caption{The distribution of $Sumdir$.
The solid line shows the histogram of supernova Monte Carlo events 
where the incident electron neutrinos from a neutronization burst
have a fixed direction. The dashed line shows the histogram of
random direction events in the SK tank.\label{sumdir}}
\end{figure}

	Since the threshold of multiplicity for a candidate cluster is two 
events, a lot of backgrounds exist even after the $R_{mean}$ cut.
Therefore, we need another variable to reduce background.
In this case, the directional information 
of observed events is a strong tool because 
the recoil electrons have almost the same direction as the incident neutrinos,
which means that events from a real neutronization burst should have 
roughly the same reconstructed direction. To check the isotropy of events 
in a candidate cluster, a new variable is defined as follows:  
\begin{equation}
Sumdir=\frac{\left| \displaystyle{\sum_{i=1}^{M}} \vec{dir_i} \right| }{M}\quad .
\end{equation}
where $\vec{dir_i}$ is a reconstructed direction vector of events in a cluster,
and $M$ is the multiplicity of the cluster . 
By this definition, $Sumdir$ will be close to 1 in the case of a 
real supernova cluster. 
Figure~\ref{sumdir} shows the $Sumdir$ distributions of 
supernova Monte Carlo and background Monte Carlo.
We set the threshold of $Sumdir$ for a supernova candidate cluster 
to 0.75 as this is the point at which the signal-to-noise ratio exceeds unity; 
the efficiency for real supernova events is estimated to be 84\%.  

	Table~\ref{tab:shtw} shows the number of observed candidates 
during the periods of SK-I and SK-II after $R_{mean}$ and $Sumdir$ cuts.
There is good agreement between the number of observed clusters
and the number of expected backgrounds which can be estimated by 
\begin{equation}
N_{bg} = \sum_{i=start}^{stop} \sum_{j=M_{thr}}^{\infty} R_i T_i
 \frac{e^{ - R_i \Delta T } \left[R_i \Delta T\right]^{j-1}}{(j-1)!} 
\end{equation}
where $M_{thr}$ is threshold of multiplicity, $R_i$ and $T_i$ are single event rate 
and live time of each data taking period (maximum 24 hours), 
and $\Delta T$ is the time-window for each case.
We can use a Poisson-based estimate for the background because, after the $R_{mean}$ cut,
most of the background clusters are due to chance coincidences of
low energy events such as solar neutrino events, 
flasher events, and spallation events in the data samples.
Since there were no candidates with stricter criteria:
$\geq$ 3 events / 1 msec, $\geq$ 3 events / 10 msec, or $\geq$ 3 events / 100 msec,
also in agreement with the expected background as shown in Table ~\ref{tab:shtw},
we conclude that no signal from a real neutronization burst
was observed during this period. 

\section{Discussion and conclusions}

	We have searched the SK-I and SK-II data
for neutrino burst signals from supernova explosions.
We conclude that no real signals of supernova bursts
occurred during the data taking periods between late May 1996 and early October 2005,
which corresponds to a total livetime of 2589.2 days
\footnote{Even during periods of calibration we found no clear galactic 
supernova neutrino bursts in the SK detector.}. 
Super-K also performed an all sky search for transient
astrophysical neutrinos in the GeV-TeV energy region and did not find
anything~\cite{astrnu}, so we can rule out the detection of high energy neutrinos from
any supernovae as well.
 
	We can evaluate the performance of 
Super-K as a supernova watcher based on these results.
We simulate neutrino events in the tank to estimate detection probability 
of a supernova as a function of distance from the earth.
The incident neutrinos are assumed to be emitted by a supernova of 
the model used by the Livermore group ~\cite{Totani}.
The detection probability of a supernova at a certain distance is determined as 
a probability in which one simulated neutrino burst satisfies each criterion
given in previous sections after basic data reduction.

\begin{figure}[h]
\epsscale{1.2}
\plotone{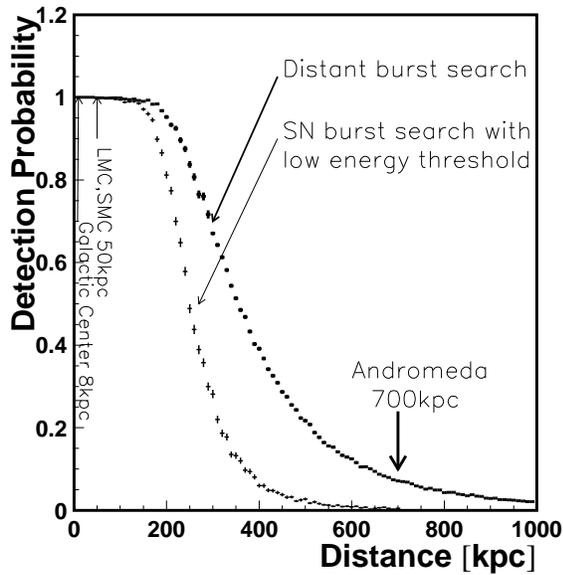}
\caption{The probability of detecting supernovae assuming
a specific supernova model at SK.
Full (100\%) detection probability is retained out to around 100 kpc.\label{dprb}}
\end{figure}

	As shown in Figure ~\ref{dprb}, full detection probability is maintained
out to around 100 kpc. Therefore, the upper limit at 
90\% C.L.  for the supernova explosion rate out to
100~kpc --- within which our Galaxy, the LMC, and the SMC may be found ---  
is determined to be 0.32 per year by combining the results from SK-I and SK-II.
While the probability for the burst search with lower energy thresholds
goes down rapidly to almost 0.0 at 700 kpc --- the distance to the Andromeda 
Galaxy --- the probability of the distant supernova search
is still 0.075 at this distance, which demonstrates the 
benefit of conducting a long time-window search in addition to the 
usual burst search. 

	We have also performed, for the first time, a systematic search for 
neutrinos from neutronization bursts.  These could occur in isolation in   
the case of early black hole formation following a core collapse. 
However there was no such signal observed in the data set
with a total livetime of 2,381.3 days.    

\section*{Acknowledgments}

	The authors gratefully acknowledge the cooperation of the Kamioka Mining
and Smelting Company. Super-K has been built and operated from
funds provided by the Japanese Ministry of Education, Culture, Sports,
Science and Technology, the U.S. Department of Energy, and the U.S.
National Science Foundation. This work was partially supported by the
Korean Research Foundation (BK21), the Korean Ministry of Science and
Technology, and the National Science Foundation of China.

\begin{deluxetable}{ccccc}
\tablewidth{0pc}
\tablecaption{Number of candidates 
and backgrounds in neutronization burst search \label{tab:shtw}}
\tablehead{
\colhead{}    &  \multicolumn{2}{c}{SK-I} &\multicolumn{2}{c}{SK-II} \\
\cline{2-3} \cline{4-5} \\
\colhead{Criterion} & \colhead{Candidate} & \colhead{BG\tablenotemark{a}}  &
\colhead{Candidate}    & \colhead{BG\tablenotemark{a}} }
\startdata
\\[-5pt]
 $\geq$ 2events/1msec   & 1   & 2.10  &     0         & 0.125 \\
 $\geq$ 2events/10msec  & 19  & 19.1  &     0         & 1.25 \\
 $\geq$ 2events/100msec & 194 & 191   &    10         & 12.5   \\
\cline{1-5}
\\[-5pt]
 $\geq$ 3events/1msec   & 0   & 9.90$\times 10^{-6}$  &    0        &  1.65$\times 10^{-7}$ \\
 $\geq$ 3events/10msec  & 0   & 9.78$\times 10^{-4}$  &    0        &  1.65$\times 10^{-5}$ \\
 $\geq$ 3events/100msec & 0   & 9.78$\times 10^{-2}$  &    0        &  1.65$\times 10^{-3}$ 
\enddata
\tablenotetext{a}{The number of backgrounds was calculated from 
the chance coincidence rate of non-supernova signals such as solar neutrino events, 
flasher events, and spallation events in the data samples.}
\end{deluxetable}


\begin{thebibliography}{99}
\bibitem[Abe et al.~2006]{astrnu} Abe,~K. {\em et al,,} 2006, Astrophys. J. {\bf 652}, 198.
\bibitem[Ahrens et al.~2002]{AMND} Ahrens,~J. {\em et al,,} 2002, Astropart. Phys. {\bf 16}, 345.
\bibitem[Alekseev et al.~1987]{BKS}  Alekseev,~E.~N. {\em et al.,} 1987, Pisma Zh. Eksp. Teor. Fiz {\bf 45}, 461. 
                                     [JETP Lett. {\bf 45}, 589 (1987)]
\bibitem[Alekseev et al.~2002]{BKS_sn}  Alekseev,~E.~N. {\em et al.,} 2002, Zh. Eksp. Teor. Fiz {\bf 95}, 10. 
                                     [2002, J. Exp. Theor. Phys {\bf 95}, 10 ]
\bibitem[Ambrosio et al.~2004]{MCR} Ambrosio,~M. {\em et al.,} 2004, Eur. Phys. J. C {\bf 37}, 265.
\bibitem[Ando et al.~2005]{ando} Ando,~S. {\em et al.,} 2005, Phys. Rev. Lett. {\bf 95}, 171101.
\bibitem[Bionta et al.~1987]{IMB}  Bionta,~R.~M. {\em et al.,} IMB collaboration, 1987, Phys. Rev. Lett. {\bf 58}, 1494.
\bibitem[Dadykin et al.~1987]{LSD}  Dadykin,~V.~L. {\em et al.,} 1987, Pisma Zh. Eksp. Teor. Fiz {\bf 45}, 464. 
                                     [1987, JETP Lett. {\bf 45}, 593 ]
\bibitem[Dye et al.~1989]{IMB_sn} Dye,~S.~T. {\em et al.,} 1989, Phys. Rev. Lett. {\bf 62}, 2069.
\bibitem[Hirata et al.~1987]{KAM} Hirata,~K.~S. {\em et al.,} 1987, Phys. Rev. Lett. {\bf 58}, 1490.
\bibitem[Hosaka et al.~2006]{SK} Hosaka,~J. {\em et al.,} 2006, Phys. Rev. D {\bf 73}, 112001 
\bibitem[Jegerlehner et al.~1996]{Jegerlehner} Jegerlehner,~B. {\em et al.,} 1996, Phys. Rev. D {\bf 54}, 1194.
\bibitem[Langanke et al.~1996]{langanke} Langanke,~K. {\em et al.,} 1996, Phys. Rev. Lett. {\bf 76}, 2629
\bibitem[LVD Coll~2003]{LVD}  LVD Coll, 2003, proc. 28th ICRC H.E 2.3.9, Tokyo. 
\bibitem[Nakahata et al.~1999]{LINAC} Nakahata,~M. {\em et al.,} 1999, Nucl. Instrum. Methods Phys. Res. Sect. A 
{\bf 421}, 113.
\bibitem[Raffelt~2002]{Raffelt} Raffelt,~G.~G. Nucl. Phys. B (Pro. Suppl) 2002, {\bf 110}, 254.
\bibitem[Sumiyoshi et al.~2005]{Sumiyoshi} Sumiyoshi,~K. {\em et al.,} 2005, Astrophys. J. {\bf 629}, 922.
\bibitem[Sumiyoshi et al.~2006]{sumi} Sumiyoshi,~K. {\em et al.,} 2006, Phys. Rev. Lett. {\bf 97}, 91101.
\bibitem[Suzuki~1993]{kam} Suzuki,~Y. 1993, {\em in Proc.of the International Symposium on Neutrino Astrophysics: 
Frontiers of Neutrino Astrophysics}, edited by Y.~Suzuki and K.~Nakamura, 
(Universal Academy Press Inc., Tokyo), number 5 in Frontier Science Series, p.61.
\bibitem[Thompson et al.~2003]{Burrows} Thompson,~T.~A. {\em et al.,} 2003, Astrophys. J. {\bf 592}, 434.
\bibitem[Totani et al.~1998]{Totani} Totani,~T. {\em et al.}, 1998, Phys. Rev. Lett. {\bf 80}, 2039.
\bibitem[Takahashi et al.~2001]{OSC} Takahashi,~K. {\em et al.,} 2001, Phys. Rev. D {\bf 64}, 93004.
\end{thebibliography}
\end{document}